\documentclass[12pt]{article}
\usepackage{graphicx}
\usepackage{amssymb}
\input epsf

\def\beq{\begin{equation}}
\def\eeq{\end{equation}}
\def\psid{\psi^{\dagger}}
\def\phid{\phi^\dagger}
\def\pd{\partial}
\def\by{\bar{y}}
\def\e{\epsilon}
\def\ep{\epsilon}

\def\lra{\leftrightarrow}
\def\bal{\bar{\alpha}}
\title{\bf Four-pomeron vertex}
\author{M.A. Braun
Dep. of High Energy physics,\\
 Saint-Petersburg State University, Russia}

\begin{document}
\maketitle
\begin{abstract}
The four-pomeron vertex is studied in the perturbative QCD. Its dominating terms  of the leading (zeroth and first) orders
in the coupling constant and subdominant in the number of colors are constructed. The vertex consists of two terms, one with a derivative in rapidity $\pd_y$
and the other with the BFKL interaction between pomerons. The corresponding part of the action
and equations of motion are found. The iterative solution of the latter is possible only for rapidities smaller than 2 and quite large coupling
constant $\alpha_s$, of the order or greater than unity, when the quadruple pomeron interaction is relatively small. Also iteration of the
part with $\pd_y$ is unstable in the infrared region and compels to introduce an infrared cut.
The variational approach with simple trying functions allows to find the minimum of the action at $\alpha_s$ of the order 0.2 and
rapidities up to 25. Numerical estimates for O-O collisions show that actually the influence of the quadruple pomeron interaction
turns out to be rather small.
\end{abstract}

\section{Introduction}
For many years the high-energy behavior in the QCD
has been the subject of intensive study both in the context of the so-called
JIMWLK approach (see e.g.~\cite{jimwlk} and references therein) and of the BFKL-Bartels
approach based on summation of the diagrams constructed for the interaction of
reggeized gluons ('reggeons') and summarized in the effective QCD reggeon
action (~\cite{lipatov}). For the interaction of a point projectile with a
large nucleus and in the approximation of a large number of colors  $N_c$
in both approaches one arrives at a simple closed equation, the Balitski-Kovchegov (BK) equation ~\cite{bal,kov},
 which actually sums the fan
pomeron diagrams constructed with the BFKL Green functions and the triple pomeron vertex.
In our papers ~\cite{braun1,braun2,braun3}
this equation has been generalized for collisions of two heavy nuclei. Unlike
the BK equation the analogous equation for AB collisions is no more an
equation for evolution in rapidity but rather the one with boundary conditions at initial and final rapidities
and so much more difficult, as illustrated by quite few attempts at its solution
~\cite{braun, bond1,bond2}.

Both the BK equation and its generalization made so far are based on the
pomeron interaction via the triple pomeron vertex. While for the BK
equation in the adopted approximation (lowest order, absence of loops) it is sufficient,
it is not so for the interaction of heavy nuclei where also interactions
mediated by the quadruple-pomeron vertex may be important. Its appearance can be
traced to the form of the gluon production in ~\cite{dusling}, which obviously
included production from the quadruple pomeron vertex. Studies in the drastically
simplified ("toy") models without transverse dimensions have shown a strong influence
of the quadruple pomeron interactions on the high-energy behavior ~\cite{mueller, braun4}.
For this reason it is interesting and important to study the quadruple pomeron vertex in the QCD,
which is the subject of this article. In the equations for the nucleus-nucleus amplitude it will appear as a new
interaction.

We recall that these equations are obtained from the non-local action
\beq
S_{AB}=S_0+S_I+S_E
\label{eq1}
\eeq
where all parts are obtained from the bilocal fields $\psi(y;r_1,r_2)$ and $\psi^\dagger(y;r_1,r_2)$
describing incoming and outgoing pomerons and depending on rapidity $y$ and two spatial points $r_1$ and $r_2$
of the two reggeons of which they consist. In Eq. (\ref{eq1}) part $S_0$ describes free pomerons
\beq
S_0=\int dyd^4zd^4z'\psi^\dagger(y,z)\Big(\pd_y+H\Big)(z,z')\psi(y,z').
\label{eq2}
\eeq
Here $z$ combines the two points $r_1$ and $r_2$: $z=\{r_1,r_2\}$.

Hamiltonian $H$ can be taken in the form symmetric
in the incoming and outgoing pomerons. Then the pomeron $P(k_1,k_2)$ ("semi-amputated")
in the momentum space is related to the standard BFKL pomeron $P_{BFKL}(k_1,k_2)$
as
\beq
 P(k_1,k_2)=k_1k_2P_{BFKL}(k_1,k_2)
\label{eq0}
\eeq
and
\[
H(k_1,k_2|k_1',k_2')=-(2\pi)^2\delta^2(k_1+k_2-k'_1-k'_2)\Big[(2\pi)^2\delta^2(k_1-k_1')\Big(\omega(k_1)+\omega(k_2)\Big)\]\beq+
\frac{2\alpha_sN_c}{k_1k_2k'_1k'_2}\Big(\frac{k_1^2{k'_2}^2+k_2^2{k'_1}^2}{(k_1-k_1')^2}-(k_1+k_2)^2\Big)\Big],
\label{eq3}
\eeq
where $\omega(k)$ is the reggeized gluon trajectory minus one. The free action defines the pomeron propagator
 $G(y;z,z')$ which satisfies
\beq
(\pd_y+H)G(y;z,z')=\delta(y)\delta^4(z-z').
\label{eq4}
\eeq
In correspondence with (\ref{eq0}) it is related to the standard BFKL Green function by
\beq
G(y;z,z')=T(z)G_{BFKL}(y;z,z')T(z'),\ \ T(z)=\sqrt{\nabla_{r_1}^2\nabla_{r_2}^2}.
\label{eq5}
\eeq

Part $S_E$ determines the interaction of the pomerons with the participant nuclei.
It is assumed that the sources for the projectile B and target A are different from
zero at rapidities $y=Y_B$ and $y=Y_A$ respectively:
\beq
S_E=-\int dy d^2z \Big(w_B(z)\psi (y,z)\delta(y-Y_B)+\psid(y,z)w_A(z)\delta(y-Y_A)\Big).
\label{sext}
\eeq
If $Y_B>Y_A$ then $\psi(y,z)=0$ at $y<Y_A$ and $\psid(y,z)=0$ at $y>Y_B$.
So in the action the integration over $y$ actually extends  over the interval
$Y_A<y<Y_B$.

The  part $S_3$ describes the interaction between the pomerons.
Both in the BK equation and its generalization for AB scattering made so far it was
taken as  the triple pomeron interaction.
Then in the configuration space it has the form
\beq
S_3=\frac{2\alpha_s^2N_c}{\pi}
\int dy\frac{d^2r_1d^2r_2d^2r_3}{r_{12}^2r_{23}^2r_{31}^2}
\Big(T^{-1}(z_1)\psi_(y,z_1\Big)\Big(T^{-1}(z_2)\psi_(y,z_2)\Big)\Big( r_{12}^4T(z_3)\psid (y,z_3)\Big) +{\rm h.c},
\eeq
where $z_1=\{r_2,r_3\}$,  $z_2=\{r_3,r_1\}$, $z_3=\{r_1,r_2\}$.
This expression looks quite complicated due to nonlocal operators $T$. However in diagrams
it is coupled to pomeron propagators, which also contain these operators so that in the end
all the non-locality is eliminated.

Our aim is to find an additional part for interaction $S_4$ which comes from the quadruple
pomeron interaction. This part contains various contributions with  different orders in
the coupling $g$ and number of colors $N_c$. So our first task will be to study these orders,
which will be done in the next section. The third section will be devoted to the derivation
of the leading contribution for $S_4$. In the fourth section we shall discuss
the changes in the coupled pomeron equations due to quadruple interaction in the no-loop approximation.
In the fifth section we shall estimate the actual contribution of this
interaction to the total O-O cross-section.
Some conclusion will be presented in the last section.

\section{Orders of magnitude}
To estimate the orders of magnitude in the  pomeron interactions
one can use the simplest diagrams with projectile and targets represented by simple quarks
attached to the lowest order pomerons (just two-gluon exchange). Additional reggeon interactions in the BFKL approach will contribute
contributions of the relative orders $(\alpha_sN_cY)^n\sim 1$ where $Y$ is the overall rapidity.
One should take into account that  diagrams with a vertex, that is with an interaction
between reggeons at a fixed rapidity $y$, will contain factor $Y$ due to integration over $y$.
In the pure perturbative approach one should additionally
take into account the coupling to external quarks. So the rule is to
take lowest order diagrams with quarks as
participants and divide by factor $(\alpha_s N_c)^{n_P}$
where $n_P$ is the number of participants.
\begin{figure}
\begin{center}
\includegraphics[width=15 cm]{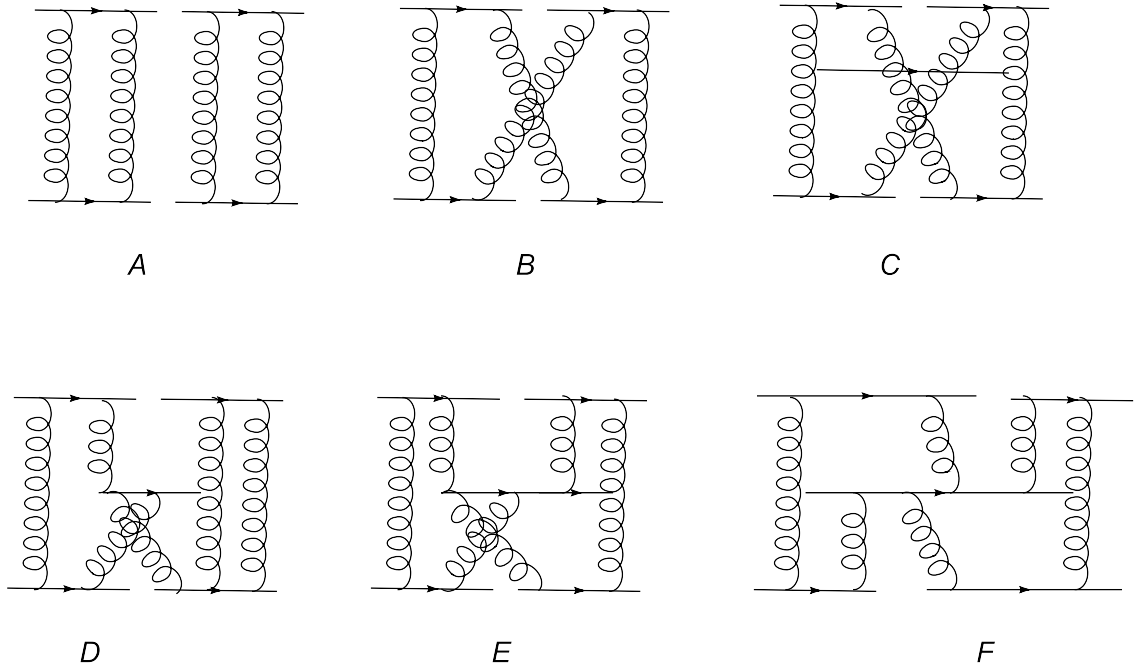}
\caption{Lowest order diagrams which involve interaction of four pomerons.}
\label{fig1}
\end{center}
\end{figure}
In this way for the triple pomeron vertex  we find the total
order $\alpha_s^5N_c^4$ and for the vertex, dividing by
$(\alpha_s N_c)^3$, we obtain the correct factor $\alpha_s^2N_c$.

The quadruple pomeron vertex,  unlike the triple one, does not change the number of reggeons.
The diagrams of the lowest order include no pomeron interaction at all.
The dominant diagram is of course the disconnected one  Fig. \ref{fig1},A
of the total order $\alpha_s^4 N_c^4$. With $N_P=4$ this gives its final order unity.
\beq
{\rm order}\,D_A=1.
\eeq
There is another diagram with no pomeron interaction but with a redistribution of color
in the collision  Fig. \ref{fig1},B. It will give a contribution to the quadruple pomeron interaction.
Its order is $\alpha_s^4 N_c^2$. This gives order $1/N_c^2$ to the resulting quadruple pomeron vertex.

Next we consider diagrams with reggeon interactions. As mentioned they will contain factor $Y$.
The lowest order diagram is shown in Fig. \ref{fig1},C.
Its order is $\alpha_s^5N_c^3Y$.
Dividing by $(\alpha_s N_c)^4$ we obtain the factor for the
corresponding  four-pomeron vertex $\alpha_s/N_c$.
Factor $Y\sim 1/\alpha_sN_c$ converts the total contribution into $1/N_c^2$,
so that effectively this diagram gives a contribution of the same order as Fig. \ref{fig1},B.
\beq
{\rm order}\,D_{B,C}=\frac{1}{N_c^2}.
\eeq

Diagrams with transition of 2 reggeons into 3 Fig. \ref{fig1},D and
3 reggeons into 3  Fig. \ref{fig1},E
 have both order $\alpha_s^6 N_c^4Y$, which is $\alpha_sN_c$ smaller than $D_{B,C}$.

 \[
 \frac{{\rm order}\,D_{D,E}}{{\rm order}\,D_{B,C}}=\alpha_sN_c<<1,
 \]
 which follows from $\alpha_sN_c y\sim 1$ at $Y>>1$ assumed in the BFKL approach.

 Finally consider the diagram with all 4 reggeons interacting  Fig. \ref{fig1},F.
 Its order is $\alpha_s^7N_c^5Y$, smaller by factor $\alpha_cN_c$ than $D_{D,E}$.
 So
 \[
 \frac{{\rm order}\,D_F}{{\rm order}\,D_{B,C}}=(\alpha_sN_c)^2
 \]

 So to conclude, the dominant contribution to the quadruple interaction comes from
 diagrams of the type  Fig. \ref{fig1},B and C. Vertices corresponding to diagrams
 Fig. \ref{fig1},D,E and  Fig. \ref{fig1},F are smaller by factors $\alpha_sN_c$
 and $(\alpha_cN_c)^2$ respectively.

The relative order of the 3-pomeron to 4-pomeron vertices is
\beq
\frac{3-pomeron}{4-pomeron}=\alpha_sN_c^2
\eeq
The BFKL approach assumes $\alpha_sN_C<<1$ but we take $N_c>>1$,
so this relative order is in fact undetermined.

\section{Quadruple pomeron vertex in the lowest approximation}
\subsection{Fig. \ref{fig1}.B}
As follows from the previous section the dominant contributions to the 4-pomeron interaction
come from the diagrams of the type shown in Figs, \ref{fig1},B and C. We start from the first diagram
Fig. \ref{fig1}.B.

\begin{figure}
\begin{center}
\includegraphics[width=6 cm]{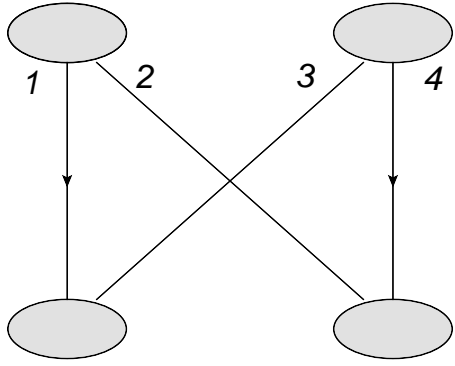}
\caption{Color redistribution of 4 pomerons}
\label{fig2}
\end{center}
\end{figure}

Taking for the participants two incoming  and two outgoing pomerons we shall actually consider the diagram shown in
Fig. \ref{fig2}.
To write down the expression for this diagram it is convenient to make use of the multirapidity formalism for the reggeon dynamics
introduced in ~\cite{braun5}. In this formalism resembling the old Gribov technique, each reggeon is assumed to possess its own "energy"
$\e$ apart from its transverse momentum $k$ with the propagator
\beq
\Delta(\ep,k)=\frac{i}{\ep+\omega(k)+i0}.
\eeq
This energy is assumed to be conjugated to the rapidity variable $y$, which is established by the transition of time $t$ to rapidity $y$
according to $it\to y$.
At each interaction both energies and transverse momenta are conserved. In this formalism the pomeron
with energies and momenta of its two component reggeons $\e_1,k_1$ and $\e_2,k_2$  can be presented as
\beq
{\cal P}(\ep_1,\ep_2;k_1,k_2)=\frac{F(E_{12};k_1,k_2)}{(\ep_1+\omega(k_1)+i0)(\ep_2+\omega(k_2)+i0)}
\label{pom1}
\eeq
where $E_{12}+\ep_1+\ep_2$ is the total pomeron energy. The "normal" pomeron $P(E_{12},k_1,k_2)$ depending on its own energy $E_{12}$
conjugated to its rapidity is given by
\beq
P(E_{12},k_1,k_2)=\frac{F(E_{12};k_1,k_2)}{E_{12}+\omega(k_1)+\omega(k_2)}.
\label{pom2}
 \eeq

 Now we turn to the diagram in Fig.\ref{fig2}. We assume that the two incoming pomerons have their energies
 $E_{12}$ and $E_{34}$ and the two outgoing pomerons $E_{13}$ and $E_{24}$. The conservation law gives factor
 \[
 2\pi \delta(E_{12}+E_{34}-E_{13}-E_{24})\]
 which will be suppressed in the following. Conservation of energy at each pomeron further restricts the reggeon energies  to a single
 energy, say, $\e_1$ with the rest determined as follows
 \beq
\ep_2=E_{12}-\ep_1,\ \
\ep_3=E_{13}-\ep_1,\ \
\ep_4=E_{24}-E_{12}+\ep_1.
\label{eps}
\eeq
The diagram also depends on fixed total momenta of the pomerons $k_{12}=k_1+k_2$ and so on
with $k_{12}+k_{34}=k_{13}+k_{24}$. As a result there is only one independent transverse momentum, say $k_1$
and the rest given as
\[ k_2=k_{12}-k_1,\ \ k_3=k_{13}-k_1,\ \ k_4=k_{34}-k_{13}+k_1.\]

 So the diagram will be given by the integral over energy $\ep_1$ and transverse momentum $k_1$.
 Suppressing the latter integration we find
\beq
D_B(E_{12},E_{34},E_{13},E_{24})=\frac{1}{N_c^2}
\int\frac{d\ep_1}{2\pi}
\frac{F_{12}F_{34}F_{13} F_{24}}
{(\ep_1+\omega_1)(\ep_2+\omega_2)(\ep_3+\omega_3)(\ep_4+\omega_4)},
\label{ampli}
\eeq
Here $F_{12}=F(E_{12};1,2)$, for brevity we denote the transverse momenta by just their numbers: $1=k_1$
and so on.
It is assumed that each $F$ carries factor $\alpha_cN_c$ which produces the overall factor $1/N_c^2$.
All $\omega$'s are assumed to have a small positive imaginary part.
Calculation of the integral over $\e_1$ gives
\[
I=\int\frac{d\ep_1}{2\pi}
\frac{1}{\ep_1+\omega_1+i0}\,
\frac{1}{E_{12}-\ep_1+\omega_2+i0}\,
\frac{1}{E_{13}-\ep_1+\omega_3+i0}\,
\frac{1}{E_{24}-E_{12}+\ep_1+\omega_4+i0}.
\]\[
=
-\frac{i}{E_{13}-E_{12}+\omega_3-\omega_2}\,
\Big\{
\frac{1}{E_{12}+\omega_1+\omega_2+i0}\,
\frac{1}{E_{24}+\omega_2+\omega_4+i0}\]\beq
-
\frac{1}{E_{13}+\omega_1+\omega_3+i0}\,
\frac{1}{E_{34}+\omega_3+\omega_4+i0}\Big\}.
\label{iint}
\eeq
 In terms of   pomerons $P(E;k_1,k_2)$ the diagram will be given after the transverse momentum integration of
\[
D_1(E_{12},E_{34},E_{13},E_{24})=
-i\frac{1}{N_c^2}\Big(\frac{1}{2}(E_{12}+E_{34}+E_{13}+E_{24})+\sum_{i=1}^4\omega_i\Big)\]
\beq\times
P_{12}P_{34}P_{13} P_{24}.
\label{ampli1}
\eeq
where  $P_{12}\equiv P(E_{12};1,2)$ and so on.
Passing to dependence on rapidities we have
\[
D_1(y_{12},y_{34},y_{13},y_{24})=\frac{1}{N_c^2}
\int \frac{dE_{12}dE_{34}dE_{13}dE_{24}}{(2\pi)^4}\]
\beq\times
2\pi\delta(E_{12}+E_{34}-E_{13}-E_{24})e^{-y_{12}E_{12}-y_{34}E_{34}+y_{13}E_{13}+y_{24}E_{24}}
D_1(E_{12},E_{34},E_{13},E_{24})
\label{dbrap}
\eeq
where we discriminate between the incoming and outgoing pomerons, $P_{12},P_{34}$ and $P_{13},P_{24}$
respectively.
We introduce the vertex rapidity $y=it$ by presenting
\[
2\pi\delta(E_{12}+E_{34}-E_{13}-E_{24})=\int dt e^{it(E_{12}+E_{34}-E_{13}-E_{24}}
\]
and for each pomeron use
\beq
\int \frac{dE}{2\pi}P(E,k_1,k_2)e^{-tE}=P(t;k_1,k_2),\ \ {\rm with}\ \  P(t;k_1,k_2)=0\ \  {\rm at}\ \  t<0
\label{ty}
\eeq
with the subsequent passage $t\to -iy$, in which an extra factor $(-i)$ appears, see \cite{braun5}).

In this way we ultimately get
\[
D_B(y_{12},y_{34},y_{13},y_{24})=\frac{1}{N_c^2}\Big(\frac{1}{2}(\pd_{y_{12}}+\pd_{y_{34}}-\pd_{y_{13}}-\pd_{y_{24}})-\sum_{i=1}^4\omega_i\Big)\]\beq
\times\int dy
P(y_{12}-y;1,2)P(y_{34}-y;3,4) P(y-y_{13};1,3)P(y-y_{24};2,4).
\label{dbi}
\eeq
Note that $P(y)$ has a jump at $y=0$.
One can  put the derivatives inside the integral and rewrite (\ref{dbi}) in the form
\[
D_B(y_{12},y_{34},y_{13},y_{24})=
\frac{1}{N_c^2}\int dy\]\beq\times
P(y_{12}-y;1,2)P(y_{34}-y;3,4)\Big(\frac{1}{2}\stackrel{\leftrightarrow}{\pd_y}-\sum_{i=1}^4\omega_i\Big) P(y-y_{13};1,3)P_{24}(y-y_{24};2,4).
\label{dbia}
\eeq
So  the contribution from the  diagram Fig. \ref{fig1},B splits into two parts
\beq
D_B^{(1)}=
\frac{1}{N_c^2}\frac{1}{2}\int dy\Big(
P(y_{12}-y;1,2)P(y_{34}-y;3,4)\stackrel{\leftrightarrow}{\pd_y}P(y-y_{13};1,3)P_{24}(y-y_{24};2,4)\Big)
\label{dbi1}
\eeq
and
\beq
D_B^{(2)}=
-\frac{1}{N_c^2}\sum_{i=1}^4\omega_i\int dy
P(y_{12}-y;1,2)P(y_{34}-y;3,4) P(y-y_{13};1,3)P_{24}(y-y_{24};2,4).
\label{dbi2}
\eeq

The term $D_B^{(2)}$ is infrared divergent due to trajectories $\omega_i$. It
has order $\alpha_sN_c$ and is accompanied by factor Y in the diagram, so it has the same
order as the diagram in Fig. \ref{fig1},C to be considered later. In the sum with the latter
the infrared divergence will be canceled.

The  term $D_1^{(1)}$ is infrared finite. To see its order  it is convenient to transform it using the BFKL equation (\ref{eq4}).
The incoming pomeron can be presented via the Green function as
\beq
P_{12}(y_{12}-y,1,2)=\int d^4K'\rho_{12}(1,2)(K')G(y_{12}-y;K',K)\equiv \rho_{12} G(y_{12}-y)
\label{pomin}
\eeq
where $K$ combines $k_1$ and $k_2$, $\rho_{12}$ is the color source and in the second equality $\rho$ and $G(y)$ are considered
as operators in the $K$ space, conjugate to the coordinate $z$-space introduced in the Introduction.
The Green  function satisfies Eq. (\ref{eq4}), which is the operatorial notation is
\beq
\pd_yG(y)=1-HG(y)=1-G(y)H.
\eeq
From this equation we find
\beq
\pd_{y_{12}}P_{12}(y_{12}-y)=\rho_{12}\delta(y_{12}-y)-P_{12}(y_{12}-y)H_{12}
\label{deriv1}
\eeq
where the momentum or coordinates are suppressed and it is assumed that Hamiltonian $H_{12}$ acts on them.
Similarly for the second incoming pomeron
\beq
\pd_{y_{34}}P_{34}(y_{34}-y)=\rho_{34}\delta(y_{34}-y)-P_{34}(y_{34}-y)H_{34}.
\label{deriv2}
\eeq

For the outgoing pomeron we have
\beq
P_{13}(y-y_{13};1,3)=\int d^4K'G(y-y_{13};K,K')\rho_(1,3)(K')\equiv G(y-y_{13})\rho_{13},
\label{pomout}
\eeq
so that we get in the same way
\beq
\pd_{y_{13}}P_{13}(y-y_{13})=-\rho_{13}\delta(y-y_{13})+H_{13}P_{13}(y-y_{13})
\label{deriv3}
\eeq
and
\beq
\pd_{y_{24}}P_{24}(y-y_{24})=-\rho_{34}\delta(y-y_{24})+H_{24}P_{24}(y-y_{24}).
\label{deriv4}
\eeq

\begin{figure}
\begin{center}
\includegraphics[width=5 cm]{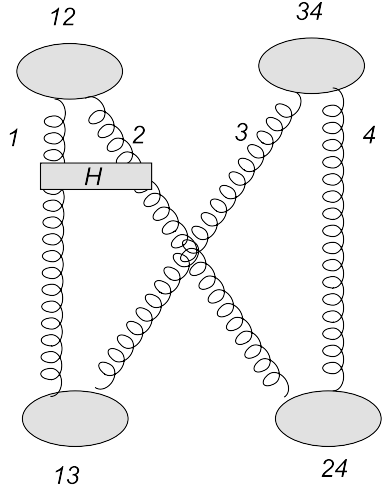}
\caption{Contribution from the interaction inside the pomerons $P_{12}$.
Three others should be added with interactions inside $P_{34}$, $P_{13}$
and $P_{24}$}
\label{fig2a}
\end{center}
\end{figure}

As a result the contribution $D_1^{(1)}$ in its turn  splits into two parts.
The first part comes from the differentiation of $\delta$ functions in rapidities:
\[
D_B^{(11)}(y_{12},y_{34},y_{13},y_{24})\]
\[ =\frac{1}{N_c^2}\int dy\Big\{\Big(\rho(1,2)\delta(y_{12}-y)
P(y_{34}-y;3,4) P(y-y_{13};1,3)P(y-y_{24};2,4)+(12)\lra (34)\Big)\]
\beq
+\Big(\rho(1,3)\delta(y-y_{13})P(y_{12}-y;1,2)P(y_{34}-y;3,4)P(y-y_{24};2,4)
+(13)\lra (24)\Big)\Big\}.
\label{dbi11}
\eeq
The second part includes the Hamiltonian $H$
\[
D_B^{(12)}(y_{12},y_{34},y_{13},y_{24})=-\frac{1}{N_c^2}\int dy P(y_{12}-y;1,2)P(y_{34}-y;3,4)\]
\beq\times
\Big(H_{12}+H_{34}+H_{13}+H_{24}\Big) P(y-y_{13}1,3)P(y-y_{24};2,4)
\label{dbi12}
\eeq
Graphically it corresponds to Fig. \ref{fig2a}.

Restoring the transverse momentum integration and introducing independent momenta for the outgoing pomerons we finally have
\[
D_B^{(1)}=
\frac{1}{2N_c^2}\int dy\int\frac{d^2k_1}{(2\pi)^2}\prod_{i=1}^4\delta^2(k_i-k'_i)\]
\beq\times
P(y_{12}-y;1,2)P(y_{34}-y;3,4)\stackrel{\leftrightarrow}{\pd_y} P(y-y_{13};1',3')P(y-y_{24};2',4')\Big)
\label{db1}
\eeq
and
\[
D_B^{(2)}=
-\frac{1}{N_c^2}\sum_{i=1}^4\omega_i\int dy\int\frac {d^2k_1}{(2\pi)^2}\prod_{i=1}^4\delta^2(k_i-k'_i)\]
\beq\times
P(y_{12}-y;1,2)P(y_{34}-y;3,4) P(y-y_{13};1',3')P(y-y_{24};2',4').
\label{db2}
\eeq

\subsection{Fig. \ref{fig1}.C}
\begin{figure}
\begin{center}
\includegraphics[width=5 cm]{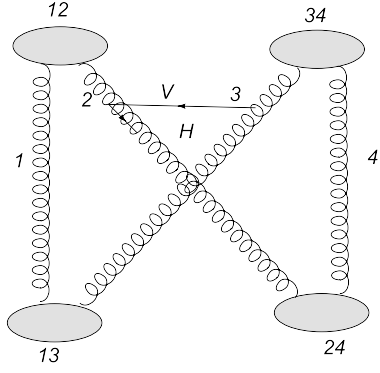}
\caption{Contribution from the interaction of reggeons 2 and 3.
The second diagram should be added with the interaction of reggeona 1 and 4}
\label{fig3}
\end{center}
\end{figure}

The diagram shown in Fig. \ref{fig1},C includes  interaction between
reggeons 2 and 3, which with the external pomerons  is shown in Fig. \ref{fig3}.
Its contribution should be summed with a similar contribution with the gluon interaction between reggeons 1 and 4.

As before we denote initial total energies as $E_{12}$ and $E_{34}$ and the final energies
as $E_{13}$ and $E_{24}$. We again suppress
the energy conservation factor $2\pi\delta(E_{12}+E_{34}-E_{13}-E_{24})$.
The diagram contains two loops and so energy integrations can be taken over $\ep_1$ and $\ep_4$.
Energies of the gluons 1, 2, 3 and 4 before the interaction are
\beq
\ep_1,\ \ E_{12}-\ep_1,\ \ E_{34}-\ep_4,\ \  \ep_4.
\label{qini}
\eeq
After the interaction gluons 2 and 3 have their energies
$E_{24}-\ep_4$ and $E_{13}-\ep_1$ respectively
and change their momenta $k_2\to k'_2$ and $k_3\to k'_3$.
Note that the  interaction connects different color configurations with
gluons from the
projectile forming colorless pairs (1,2) and (3,4) and from the target
forming colorless pairs (1,3) and (2,4). As a result the interaction has the opposite sign
as compared to the one inside the pomeron ~\cite{braun5}. If the pomeron Hamiltonian is
\beq
H_{12}=-\omega_1-\omega_2 -V
\label{hampom}
\eeq
then the interaction in Fig. \ref{fig3} is just $-V$.

So
in terms of amputated pomerons $F$ the contribution from Fig. \ref{fig3}  before the transverse integrations is
\[
D_C=i\frac{1}{N_c^2}\int\frac{d\ep_1d\ep_4}{(2\pi)^2}\]\beq\times\frac
{F_{12}F_{34})F_{13'}F_{2'4}V(2,3|2',3')}
{(\ep_1+\omega_1)(\ep_4+\omega_4)
(E_{12}-\ep_1+\omega_2)(E_{34}-\ep_4+\omega_3)
(E_{13}-\ep_1+\omega_{3'})(E_{24}-\ep_4+\omega_{2'})}.
\eeq

Integration over energies factorizes into two integrals:
\[
I_1=\int \frac{d\ep_1}{2\pi}\frac{1}
{(\ep_1+\omega_1+i0)(E_{12}-\ep_1+\omega_2+i0)(E_{13}-\ep_1+\omega_[3']+i0)}\]\beq
=-i\frac{1}{(E_{12}+\omega_1+\omega_2)(E_{13}+\omega_{3'})}
\eeq
and a similar integral over $\ep_4$ which gives
\beq
I_2=-i\frac{1}{(E_{34}+\omega_3+\omega_4)(E_{24}+\omega_{2'}+\omega_4)}.
\eeq
So passing to the pomerons we get
\beq
D_C=-i\frac{1}{N_c^2}V(2,3|2',3')
P(E_{12};1,2)P(E_{34};3,4))P(E_{13};1,3')P(E_{24};2',4).
\label{aa}
\eeq

The second amplitude $\tilde{D}_C$ with the interaction between the gluons 1 and 4 is
\beq
\tilde{D}_C=-i\frac{1}{N_c^2}V(1,4|1',4')
P(E_{12};1,2)P(E_{34};3,4))P(E_{13};1',3)P(E_{24};2,4')
\label{ab}
\eeq

Both $D_{C}$ and $\tilde{D}_{C}$ are infrared divergent. Using the fact that the BFKL Hamiltonian is infrared stable as a whole,
we can separate the divergence into the reggeon trajectories, presenting (see (\ref{hampom}))
\[
V(2,3|2',3')=-H_{23}(2,3|2',3')-(2\pi)^4\delta(2-2')\delta(3-3')(\omega_2+\omega_3)\]
and
\[
V(1,4|1',4')=-H_{14}(1,4|1',4')-(2\pi)^4\delta(1-1')\delta(4-4')(\omega_1+\omega_4).\]
Note that for arbitrary momenta of the outgoing pomerons in the transversal integral $D_{C}$ has to be multiplied
by $(2\pi)^4\delta^2(1-1')\delta^2(4-4')$ and  $\tilde{D}_{C}$ has to be multiplied
by $(2\pi)^4\delta^2(2-2')\delta^2(3-3')$. In both cases terms with $\omega_i$ acquire the product of
all 4 delta-functions between initial and final momenta. Then one finds that the terms
with delta-functions cancel with the similar terms in $D_B^{(2)}$, Eq. (\ref{db2}).
So only the terms with the BFKL Hamiltonian are left and the infrared divergence disappears.

\begin{figure}
\begin{center}
\includegraphics[width=7.2 cm]{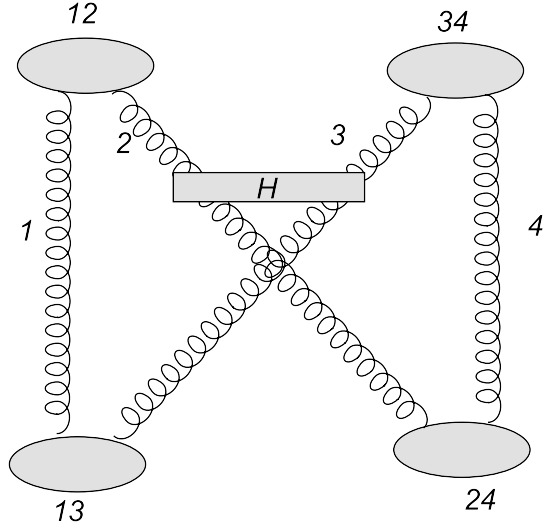}
\caption{Infrared stable contribution from the interaction between reggeons @ and 3.
The second diagram should be added with the interaction of reggeons 1 and 4}
\label{fig3a}
\end{center}
\end{figure}

After this cancelation the resulting contribution from the diagrams with one reggeon interaction can  be written as
\[
D_1=D_C+\tilde{D}_C=i\frac{1}{N_c^2}
P(E_{12};1,2)P(E_{34};3,4)P(E_{13};1',3')P(E_{24};2',4')\]\beq\times
\Big(H(2,3|2',3')(2\pi)^4\delta^2(1-1')\delta^2(4-4')+H(1,4|1',4')(2\pi)^4\delta^2(2-2')\delta^(3-3')\Big).
\label{d10}
\eeq
Transition to rapidities, as before, will give
\[
D_1(y_{12},y_{34},y_{13},y_{14})\]\[=-\frac{1}{N_c^2}\int dy
P(y_{12}-y;1,2)P(y_{34}-y;3,4)P(y-y_{13};1',3')P(y-y_{24};2',4')\]\beq\times
\Big(H(2,3|2',3')(2\pi)^4\delta^2(1-1')\delta^2(4-4')+H(1,4|1',4')(2\pi)^4\delta^2(2-2')\delta^2(3-3')\Big).
\label{d1}
\eeq
Symbolically it can be written as
\[
D_1(y_{12},y_{34},y_{13},y_{14})\]\beq=-\frac{1}{N_c^2}\int dy
P(y_{12}-y;1,2)P(y_{34}-y;3,4)\Big(H_{23}+H_{14}\Big)P(y-y_{13};1',3')P(y-y_{24};2',4')
\label{d11}
\eeq
where the relation between different momenta can be seen from Fig. \ref{fig3a}.

\section{The quadruple pomeron vertex and action}
\begin{figure}
\begin{center}
\includegraphics[width=5 cm]{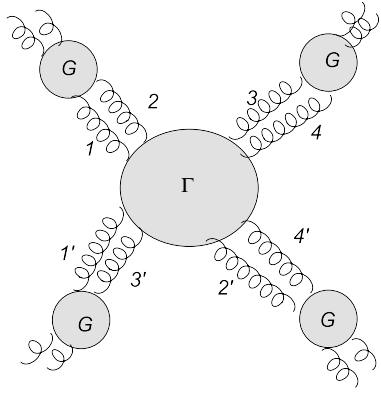}
\caption{Four pomerons coupled with the quadruple vertex}
\label{fig4}
\end{center}
\end{figure}

By definition the quadruple pomeron vertex $\Gamma$ corresponds to a contribution generated by the
coupling of four pomeron propagators, that is the Green function $G$, at rapidity $y$
of the structure
\[
\int\prod_{i=1}^4\frac{d^2k_id^2k'_i}{(2\pi)^4}G(y_{12}-y;q_1,q_2;k_1,k_2)G(y_{34}-y;q_3,q_4;k_3,k_4)(2\pi)^2\delta\Big(\sum_{i=1}^4(k_i-k'_i)\Big)\]\beq\times
\Gamma(k_1,k_2;k_3k_4|k'_1,k'_3;k'_2,k'_4)
G(y-y_{13};k'_1,k'_3;q'_1,q'_3)G(y-y_{24};k'_2,k'_4;q'_2,q'_4)
\label{defgam}
\eeq
illustrated in Fig. \ref{fig4}. In the vertex pairs of momenta are grouped to correspond to
colorless states. We also define it with the extracted $\delta$ function corresponding to  conservation f the total momentum.
Inspecting our results in the previous section we can identify contributions
$D_B^{(1)}$ and $D_1$ with this structure. In the sum they give the vertex
\[
\Gamma(k_1,k_2;k_3k_4|k'_1,k'_3;k'_2,k'_4)=\frac{1}{N_c^2}\Big\{\frac{1}{2}\stackrel{\leftrightarrow}{\pd_y}\prod_{i=1}^4(2\pi)^2\delta^2(i-i')\]\beq
-H(2,3|2',3')(2\pi)^4\delta^2(1-1')\delta^2(4-4')-
H(1,4|1',4')(2\pi)^4\delta^2(2-2')\delta^2(3-3')\Big\}.
\label{vertex}
\eeq

In accordance with (\ref{vertex}) the effective action of pomeron fields acquires new interaction terms
\beq
S_I^{(41)}=\frac{1}{2N_c^2}\int dy\prod_{i=1}^4\frac{d^2k_i}{(2\pi)^2}
\Big\{
\psid(y,k_1,k_2)\psid(y,k_3,k_4)\stackrel{\leftrightarrow}{\pd_y}\psi(y,k_1,k_3)\psi(y,k_2,k_4)\Big\}
\label{s41g}
\eeq
(note that the derivative acts on all field variables in (\ref{s41g})).
and
\[
S_I^{(42)}=\frac{1}{2N_c^2}\int dy\int\prod_{i=1}^4\frac{d^2k_id^2k'_i}{(2\pi)^4}\]\[
\Big(H(2,3|2',3')(2\pi)^4\delta^2(1-1')\delta^2(4-4')+
H(1,4|1',4')(2\pi)^4\delta^2(2-2')\delta^2(3-3')\Big)\]\beq\times\psi(y,k_1,k_2)\psi(y,k_3,k_4)\psid(y,k'_1,k'_3)\psid(y,k'_2,k'_4)
\label{s42g}
\eeq

One can write down the quasi-classical equations following from $\delta S/\delta\psi=\delta S/\delta \psid=0$.
To do this in the coordinate space one has to Fourier transform $\psi$ and $\psid$ in (\ref{s41}) taking into account also the
corresponding transformation of $H_{23}$ and $H_{14}$. However these general equations have little sense since in applications to the
nucleus-nucleus  collisions the pomerons have zero total transverse momenta. In this case the equations substantially simplify (see
~\cite{braun1,braun}) and will be studied in the next section. The general, nonforward form of (\ref{s41}), whether in the coordinate or momentum space,
is in fact essential only for the study of conformal invariance and calculating loops, the latter task remaining still very remote in future.

\section{Forward case}
In applications to the nucleus-nucleus scattering and in absence of loops all interacting pomerons have  their total transverse momentum equal to zero.
So the pomeron wave functions depend on the single transverse variable $r$ or $k$ in the coordinate or momentum spaces. For the forward pomerons it is
natural to pass to the so-called semi-amputated pomerons defining in the momentum space
\[ \psi(y,k)=k^2\phi(y,k),\ \ \psid(y,k)=k^2\phid(y,k).\]
In terms of these pomeron fields the action with only the triple-pomeron interaction was introduced in ~\cite{braun1,braun}.
\beq
S=S_0+S_3+S_E.
\eeq
The free action is then
\beq
S_0=\int \frac{dyd^2k}{(2\pi)^2}\phid(y,k)K(\stackrel{\leftrightarrow}{\pd_y}+H)\phi(y,k)
\label{s0a}
\eeq
where $H$ is the forward Hamiltonian (\ref{eq3}).
\beq
H(k,-k|k',-k')=-(2\pi)^2\delta^2(k-k')2\omega(k)-
4\alpha_sN_c\frac{1}{(k-k')^2},
\label{eq3a}
\eeq
Momenta $k$ and $k'$ are two-dimensional transversal
and $K$ is a differential operator in $k$ commuting with $H$
\beq
K=\nabla_k^2k^4\nabla_k^2.
\eeq
Appearance of the operator $K$ in (\ref{s0a}) as compared to (\ref{eq2}) is due to the transition of
the semi-amputated pomerons in the coordinate space to the momentum space.

The interaction part of the action describes splitting and merging of pomerons:
\beq
S_I^{(3)}=\frac{4\alpha_s^2N_c}{\pi}
\int\frac{dy d^2k}{(2\pi)^2}\Big({\phid}^2(y,k)K\phi(y,k)+\phi^2(y,k)K\phid(y,k)\Big).
\label{s3a}
\eeq

The external action is
\beq
S_E=-\int\frac{dyd^2k}{(2\pi)^2}\Big(w_B(k,)\phi(y,k)\delta(y-Y)+w_A(k)\phid(y,k))\delta(y)\Big),
\label{sexta}
\eeq
where $w_{A,B}$ as before describe the interaction of the pomerons with the projectile and target.
At fixed transverse point $b_A$ in the nucleus A coupling $w_A(k)$ is supposed to be proportional
to the profile nuclear function $T_A(b_A)$ and its dependence on $k$ is determined by the gluon distribution in the nucleon.

Fields $\phi$ and $\phid$ as well as $w$ are dimensionless. At fixed $b_A$ and $b_B$ action $S$ is dimensionful:
dim$\,S(b_A,b_B)=2$. It becomes dimensionless after integration over $b_A$ at fixed impact parameter $b$, so that $b_B=b_A-b$.

Note that in (\ref{s0a}) it is assumed that the integration in $y$ goes over the whole axis: $-\infty<y<+\infty$
with the conditions that $\phi(y,k)=0$ at $y<y_{min}+0$ and $\phid(y,k)=0$ at $y>y_max=Y$. At $y=0$ and $y=Y$ the piece
with $\pd_y$ contributes terms proportional to $\delta(y)$ and $\delta(y-Y)$ due to jumps of $\phi$ and $\phid$.

Turning to the quadruple pomeron interaction we have the same diagrams shown in Fig. \ref{fig1} B and C
which after the cancelation of the terms with $\omega$ give the contributions (\ref{db1}) and (\ref{d1}).
Taking $y_{12}=y_{34}=Y$ and $y_{13}=y_{24}=0$ we find
for the forward direction
\beq
D_B^{(1)}=
\frac{1}{2N_c^2}\int dy\int\frac{d^2k}{(2\pi)^2}
P_u(Y-y,k)P_u(Y-y,k)\stackrel{\leftrightarrow}{\pd_y} P_l(y,k)P_l (y,k)
\label{db1a}
\eeq
and
\beq
D_1=\frac{2}{N_c^2}\int dy\frac{d^2k d^2k'}{(2pi)^4}
P_u(Y-y,k)P_u(Y-y,k')P_l(y,k)P_l(y,k')
H(k,k'|k',k).
\label{d1a}
\eeq
For clarity we denote the upper and lower pomerons in our figures by subindexes $u$ and $l$
Note that in (\ref{d1a}) $H$ is not the forward Hamiltonian:
\beq
H(k,k'|k',k)=-(2\pi)^2\delta^2(k-k')2\omega(k)-
\frac{2\alpha_sN_c}{k^2{k'}^2}\Big(\frac{k^4+{k'}^4}{(k-k')^2}-(k+k')^2\Big).
\label{eq3b}
\eeq
From these expressions we can read the corresponding quadruple pomeron vertex and action $S_4$.
From $D_B^{(1)}$ we find the vertex
\beq
\Gamma_4^{(1)}=\frac{1}{2N_c^2})\stackrel{\leftrightarrow}{\pd_y}
\label{gam1}
\eeq
with the contribution to the action
\beq
S_4^{(1)}=
\frac{1}{2N_c^2}\int dy\int\frac{d^2k}{(2\pi)^2}{\phid}^2(y,k)\stackrel{\leftrightarrow}{\pd_y}{\phi}^2(y,k)
=\frac{1}{N_c^2}\int dy\int\frac{d^2k_1}{(2\pi)^2}({\phid}^2\phi \pd_y\phi-\phi^2\phid \pd_y\phid).
\label{s41}
\eeq
For  the equations of motion, which for $\phi(y,k)$ is obtained after differentiation $\delta/\delta \phid$,
one gets a contribution
\beq
\frac{\delta}{\delta\phid}S_4^{(1)}=
\frac{2}{N_c^2}\phid \pd_y\phi^2=\frac{4}{N_c^2}\phid\phi\pd_y\phi
\label{term41}
\eeq

The second contribution to the action from the quadruple pomeron interaction
comes from (\ref{d1a})
\beq
S_4^{(2)}=\frac{2}{N_c^2}\int dy\frac{d^2k d^2k'}{(2\pi)^4}
\phi(y,k)\phi(y,k')\phid(y,k)\phid(y,k')H(k,k'|k',k).
\label{s42a}
\eeq

The classical equations of motion which follow, multiplied  by $(1/2)K^{-1}$
from the left, are
\[
\left(\frac{\partial}{\partial y}+H\right)\phi(y,k)+\frac{2\alpha_s^2N_c}{\pi}
\Big(\phi^2(y,k)+2K^{-1}\phid(y,k) K\phi(y,k)\Big)\]\[
+\frac{2}{N_c^2}K^{-1}\int\frac{ d^2k'}{(2\pi)^2}
\phi(y,k)\phid(y,k')H(k,k'|k',k)\phi(y,k')\]\beq+\frac{2}{N_c^2}K^{-1}(\phid\phi\pd_y\phi)
=\frac{1}{2}K^{-1}w_A(k)\delta(y)
\label{cleq1}
\eeq
and
\[
\left(-\frac{\partial}{\partial y}+H\right)\phid(y,k)+\frac{2\alpha_s^2N_c}{\pi}
\Big({\phid}^2(y,k)+2K^{-1}\phi(y,k)  K\phid(y,k)\Big)\]\[
+\frac{2}{N_c^2}K^{-1}\int\frac{ dk'}{(2\pi)^2}
\phid(y,k)\phid(y,k')H(k,k'|k',k)\phi(y,k')\]\beq-\frac{2}{N_c^2}K^{-1}(\phi\phid\pd_y\phid)
=\frac{1}{2}K^{-1}w_B(k)\delta(y-Y).
\label{cleq2}
\eeq

From the $\delta$-like dependence on $y$ of the external sources it follows
that the equations can be taken homogeneous in the interval $0<y<Y$,
action of the external sources substituted by the boundary conditions in rapidities
\[
\phi(0,k)+\frac{1}{N_c^2}K^{-1}(\phid(0)\phi^2(0))=\frac{1}{2}K^{-1}w_A(k),\]\beq
\phid(Y,k)+\frac{1}{N_c^2}K^{-1}(\phi(Y){\phid}^2(Y))=\frac{1}{2}K^{-1}w_B(k).
\label{bcond}
\eeq

We noted in our previous papers that already with $S_3$ the equations of motion are no more
pure evolution equations, since the initial conditions for $\phi(y)$ and $\phid(y)$ had to be
given at different rapidities, $y=0$ for $\phi$ and $y=Y$ for $\phid$. Inclusion of quadruple action
$S_4^{(1)}$ further complicates the situation, since according to (\ref{bcond}) the initial conditions now
each contain  both $\phi$ and $\phid$.

To see the orders of magnitude of different terms in these equations it is instructive to rescale
the pomeron fields $\phi$ and $\phid$,  Hamiltonian and rapidity as
\beq\phi\to\frac{\phi}{2\alpha_s},\ \ \phid\to\frac{\phid}{2\alpha_s},\ \
H=\bar{\alpha}\bar{H},\ \
y=\frac{\by}{\bar{\alpha}}
\label{rescale}
\eeq
where $\bar{\alpha}=\alpha_sN_c/\pi$
Then the equations take the form
\beq
\frac{\partial\phi(\by,k)}{\partial \by}=F\{\phi,\phid,\pd_y\phi\}
\label{cleq1n}
\eeq
and
\beq
-\frac{\partial\phi(\by,k)}{\partial \by}=F^{\dagger}\{\phi,\phid,\pd_y\phid\},
\label{cleq2n}
\eeq
where
\[
F=-\bar{H}\phi(\by,k)-
\phi^2(\by,k)-2K^{-1}\phid(\by,k) K\phi(\by,k)\]\beq
-\frac{1}{2(\bar{\alpha}\pi)^2}K^{-1}\int \frac{d^2k'}{(2\pi)^2}
\phi(\by,k)\phid(\by,k')\bar{H}(k,k'|k',k)\phi(\by,k')
-\frac{1}{2(\bar{\alpha}\pi)^2}K^{-1}(\phid\phi\pd_y\phi)
\label{cleq1m}
\eeq
and
\[
F^{\dagger}=-\bar{H}\phid(\by,k)-
{\phid}^2(\by,k)-2K^{-1}\phi(\by,k) K\phid(\by,k)\]\beq
-\frac{1}{2(\bar{\alpha}\pi)^2}K^{-1}\int \frac{d^2k'}{(2\pi)^2}
\phid(\by,k)\phi(\by,k')\bar{H}(k,k'|k',k)\phid(\by,k')
+\frac{1}{2(\bar{\alpha}\pi)^2}K^{-1}(\phi\phid\pd_y\phid).
\label{cleq2m}
\eeq
We observe that the on the right-hand side the free and triple interactions have the same order unity
(actually $\bar{\alpha}$ in the original rapidity, which corresponds to  the BK equation). The quadruple interaction has the  orders $1/\bar{\alpha}^2$,
which may be smaller or  larger as compared with the two first terms
depending on the relation between the small $\alpha_s$ and large $N_c$.

\section{Cross-sections}
At fixed overall impact parameter $b$ and rapidity $Y$ the
 nucleus-A-nucleus-B total
cross-section is given by
\beq \sigma(Y,b)=2\left(1-e^{-T(Y,b)}\right).
\eeq
In the perturbative QCD
the eikonal function $T$ is given by the sum of all connected diagrams constructed of BFKL pomerons, which interact between themselves and with the participant nuclei,
so that $T$ is just the action $S$ for fixed $Y$ and $b$
\beq
T(Y,b)=-\int d_2b_Ad^2b_B\delta^2(b_A-b_B-b)S(Y,b_A,b_B)
\label{eik}
\eeq
In the quasi-classical approximation (without loops)
action $S$ is to be calculated from the sum of (\ref{s0a}), (\ref{s3a}), \ref{s41}), (\ref{s42a}) and(\ref{sexta}) with the
solutions of evolution Eqs. (\ref{cleq1}) and (\ref{cleq2}), $\phi_{cl}(y,k)$ and $\phid_{cl}(y,k)$.

Using the equations of motion one can somewhat simplify the expression for $S$.
Indeed multiplying the first equation by $2K\phid$, the second one by $2K\phi$,
integrating both over $y$ and $k$ and summing the results one obtains a relation
\beq
2S_0+3S_3+4S_4+S_E=0,
\label{rels}
\eeq
which is valid for the classical action, that is, calculated with the
solutions of Eqs. (\ref{cleq1}) and (\ref{cleq2}). Note that this condition is fulfilled for any form
of interactions in $S_2$, $S_3$  $S_4$ and $S_E$, since it depends only on the number of field operators
in them and hermiticity.
Using this relation we can exclude, say, $S_4$
from (12) to find
\beq
S=\frac{1}{4}\Big(2S_0\{\phi_{cl},\phid_{cl}\}+S_I^{(3)}\{\phi_{cl},\phid_{cl}\}+
3S_E\{\phi_{cl},\phid_{cl}\}\Big).
\label{clt1}
\eeq

\section{Numerical estimates}
As mentioned in our old paper ~\cite{braun} one can try to find the solution of our problem by two methods.
One of them is to solve Eqs. (\ref{cleq1n}) and (\ref{cleq2n}) by iterations.  In the simple case when by symmetry
$\phid(y,k)=\phi(Y-y,k)$ in the first iteration one starts from Eq. (\ref{cleq1n})
and solves it for $\phi(y)$ choosing for $\phi(Y-y)$ some simple initial value, typically just zero. Since in this case both
the crossed term $2K^{-1}\phid(y,k) K\phi(y,k)$ and the term coming from $S_4$ are both equal to zero, the equation
turns out to be the standard BK equation and its solution is just the sum of fan diagrams attached to the projectile.
However at the next step
one finds non-zero $\phid(y)$ as $\phi(Y-y)$ from the obtained solution, so that both terms coming from $S_3$ and $S_4$ become present.
Now on the right-hand side there appears a term which itself contains $\pd_y\phi$, so that the equation is
\beq
\pd_y\phi=F(\phid,\phi,\pd_y\phi).
\label{equation}
\eeq
To solve it we divide the interval $0<y<Y$ in steps $\Delta$
and find at each step
\beq
\pd_y\phi(y+\Delta)=F\Big(\phid(y),\phi(y),\pd_y\phi(y)\Big),
\label{equation1}
\eeq
that is, using $\pd_y\phi$ obtained at the previous step on the right-hand side.
One expects that with a sufficiently small $\Delta$ the derivative $\pd_y\phi$ will not significantly change
between steps and one will have good convergence.
Solving in this way the equation one once again determines $\phi(y)$ and $\phid(y)=\phi(Y-y)$ and starts
 the third iteration of
Eq. (\ref{cleq1n}) and so on.
If the iteration procedure is convergent one determines the solution $\phi(y,k)$ and $\phid(y.k)=\phi(Y-y,k)$ together with their derivatives in rapidity.
Then one can find the external couplings $w_A$ and $w_B$ from (\ref{bcond}). So in this approach these couplings are not fixed from the start but
found after the equations for $\phi$ and $\phid$ are solved. This circumstance has to be taken into account comparing these results with the
pure perturbational calculations in Sec. 3.

Unfortunately, as was  mentioned in  ~\cite{braun} this method has a very narrow region of convergence in $A$ and $Y$, which
shrinks with the growth of atomic number and rapidity. Without the quadruple interaction for O-O collisions it converges up to rescaled rapidity $\by=2$,
which for $\bal=0.2$ corresponds to rather small actual rapidities less than 5. In attempt to consider higher rapidities and atomic
number  in ~\cite{braun} we recurred to another method trying to find the minimum of the action by a direct variational
approach choosing some trying functions for $\phi$ and $\phid$. Even with  very primitive trying functions
\beq
\phi(y=0,k)e^{\Delta y},\ \  \phid(y=Y,k)e^{\Delta(Y-y)}
\label{tryf}
\eeq
with a varying $\Delta$ in  ~\cite{braun} we could find
a minimum for the action for very large atomic numbers and rapidities, which lead to reasonable results for the eikonal function
and cross-section.  Comparison with the exact minimum found by the solution of the equation of
motion at rapidities where such solution could be found by interactions showed that the precision of
such variational approach was about 20\%.

In the present calculations our primary task was to see the influence of  inclusion of the quadruple pomeron interaction. $S_4$.
We considered both methods, the solution of the quasi-classical equations Eqs. (\ref{cleq1n}) and (\ref{cleq2n}) by iterations and
the direct variational approach.
In both approaches we used the initial wave function which gave the best results in our old iterational solution of (\ref{cleq1n}) and (\ref{cleq2n})
\beq
\phi(y=0,k)=\frac{1}{2}a\ln\Big(1+\frac{0.2181}{k^2}\Big),\ \ a=A\sigma_0T_A(b_A),
\eeq
where $\sigma_0=20.8$ mb and $k$ is in GeV/c. This function is very similar in its behavior to the often used Golec-Biernat distribution but is simpler
and leads to somewhat larger region of convergence. One can find some details of the calculations of different pieces of the action
in Appendix.

In the iterative procedure to solve (\ref{cleq1n}) and (\ref{cleq2n}) we were confronted with a new problem related to the interaction
containing the derivative $\pd_y$ from $S_4^{(1)}$. Its contribution was found to lead to very high values of $\phi(y,k)$ in the infrared region,
which prevented iterations already at very small values of $y$. Note that this problem does not exist for the second quadruple interaction coming from $S_4^{(2)}$
which admits iterations at all $k$. To overcome this difficulty we had to introduce an infrared cut at the minimal value $k_{min}=\exp\, t_{min}$
for which iterations were possible. It turned out that $t_{min}\simeq -5$.
Even with this limitation the iteration procedure  was found to be convergent practically in the same region as
without $S_4$ only when the quadruple interaction was taken very small. Its relative magnitude is determined by factor $1/\bal^2$.
For O-O collisions at the overall rapidity $Y =2$ the iterative procedure turned out to be convergent only for quite large values of the coupling constant
$\bal\geq 1$. In our calculations we took the lowest possible value $\bal=1.1$.

\begin{figure}
\begin{center}
\includegraphics[width=8 cm]{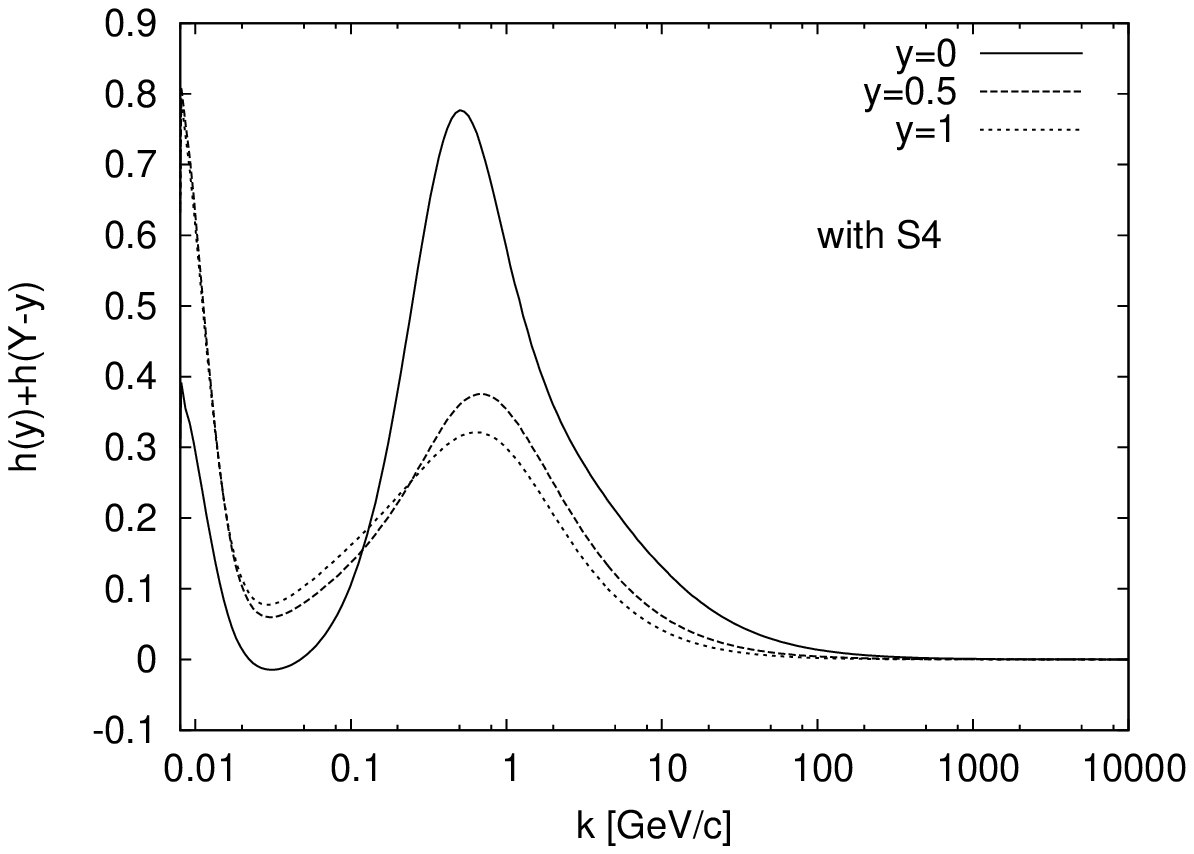}
\includegraphics[width=8 cm]{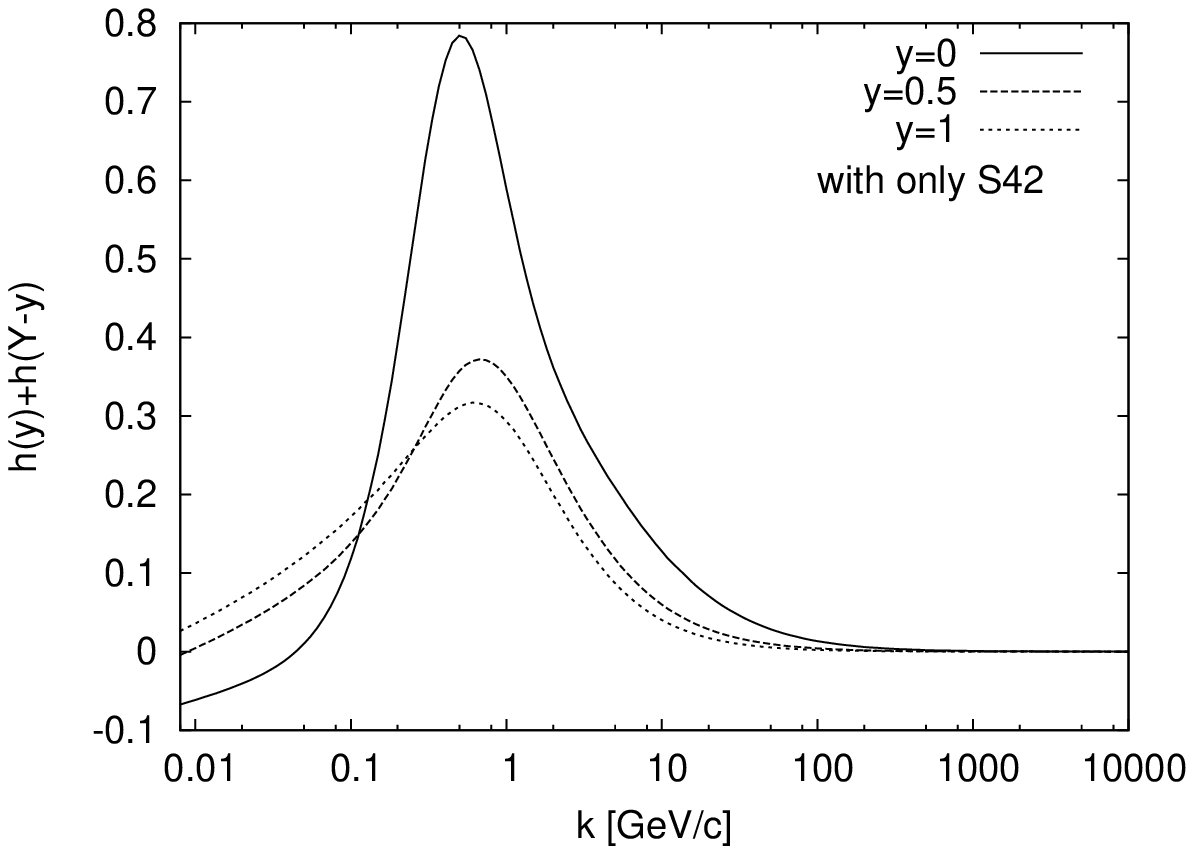}
\includegraphics[width=8 cm]{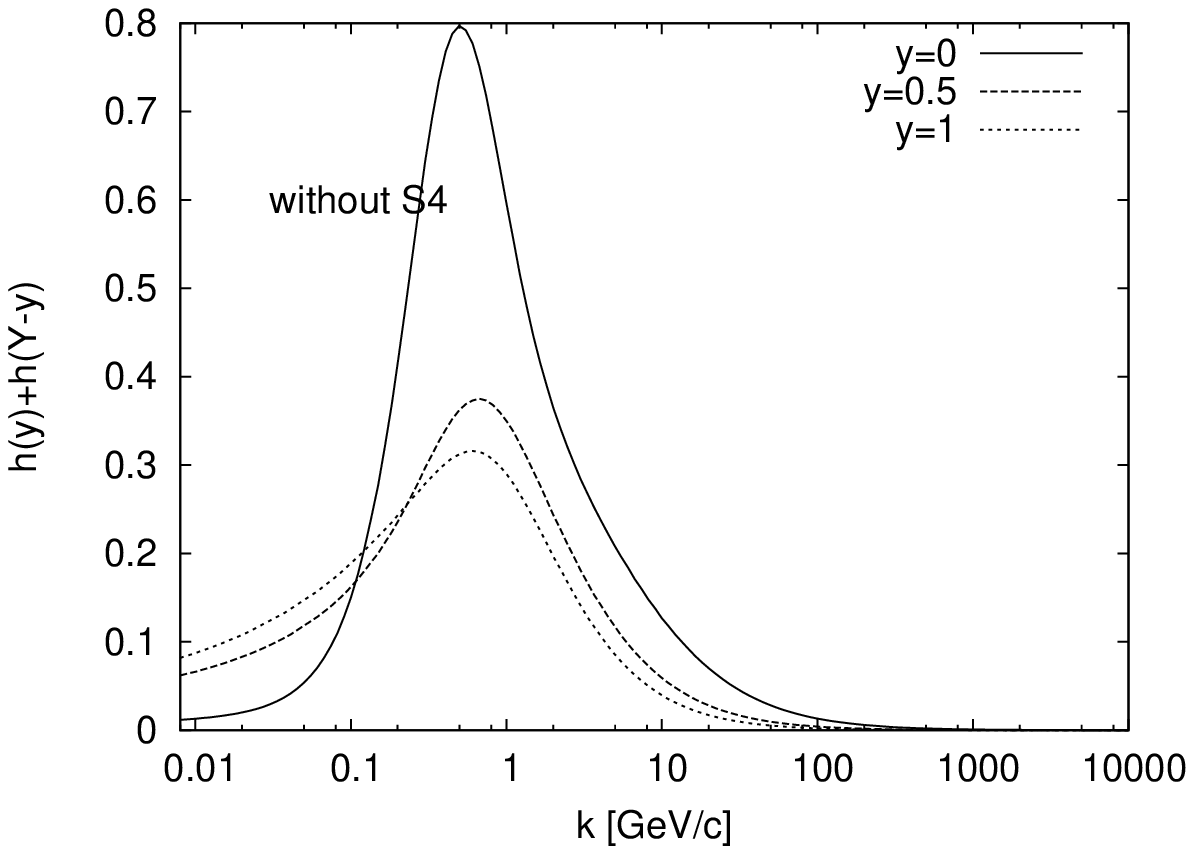}
\caption{Gluon density for O-O collisions at $Y=2$ and $\bal=1$ with the total quadruple interaction $S_4$ (upper left panel),
with only $S_4^{(2)}$ and without infrared cut (upper right panel) and
 without the quadruple pomeron interaction (lower panel)}
\label{fig5}
\end{center}
\end{figure}

The gluon density at intermediate values of rapidity $0<y<Y$ coming from both colliding nuclei can be determined as ~\cite{braun}
\beq
\frac{dxG(x,k)}{d^2bd^2k}=\frac{N_c^2}{2\pi^3\bal}\Big(h(y,k)+h^\dagger(y,k)\Big),\ \ h(y,k)=k^2\nabla_k^2\phi(y,k).
\eeq
In our actual calculations we take $b_A=b_B=0$ and $h^\dagger(y,k)=h(Y-y,k)$.
In Fig \ref{fig5} in the upper left panel we show the sum $h(y,k)+h(Y-y,k)$ for $Y=2$, $\bal=1.1$ and $y=0, 1/2$ and 1
obtained after converging iterations with the full quadruple interaction $S_4$ with the infrared cut. In  the upper right  panel
we show the same sum  with only $S^{(2)}_4$ taken into account (and without infrared cut).
To compare, in the lower panel we show the same quantity obtained by iterations without $S_4$.

\begin{figure}
\begin{center}
\includegraphics[width=8 cm]{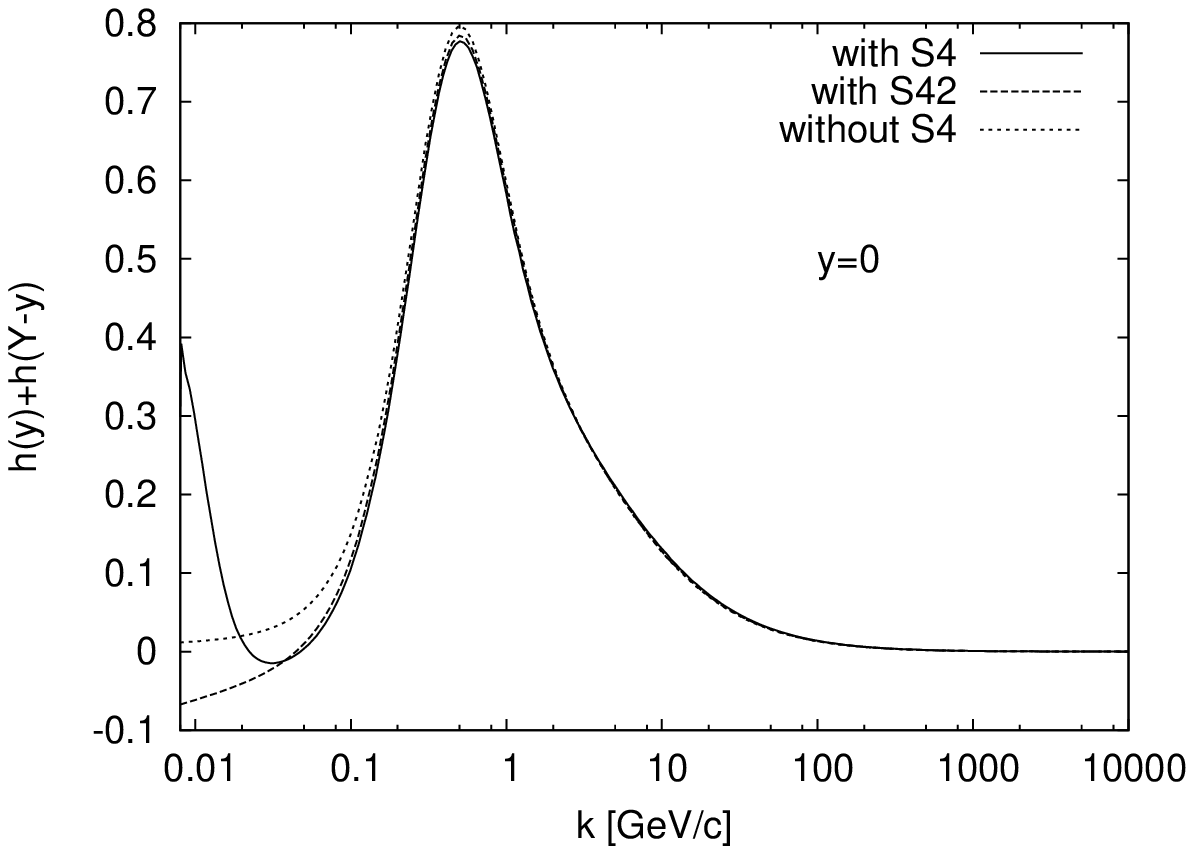}
\includegraphics[width=8 cm]{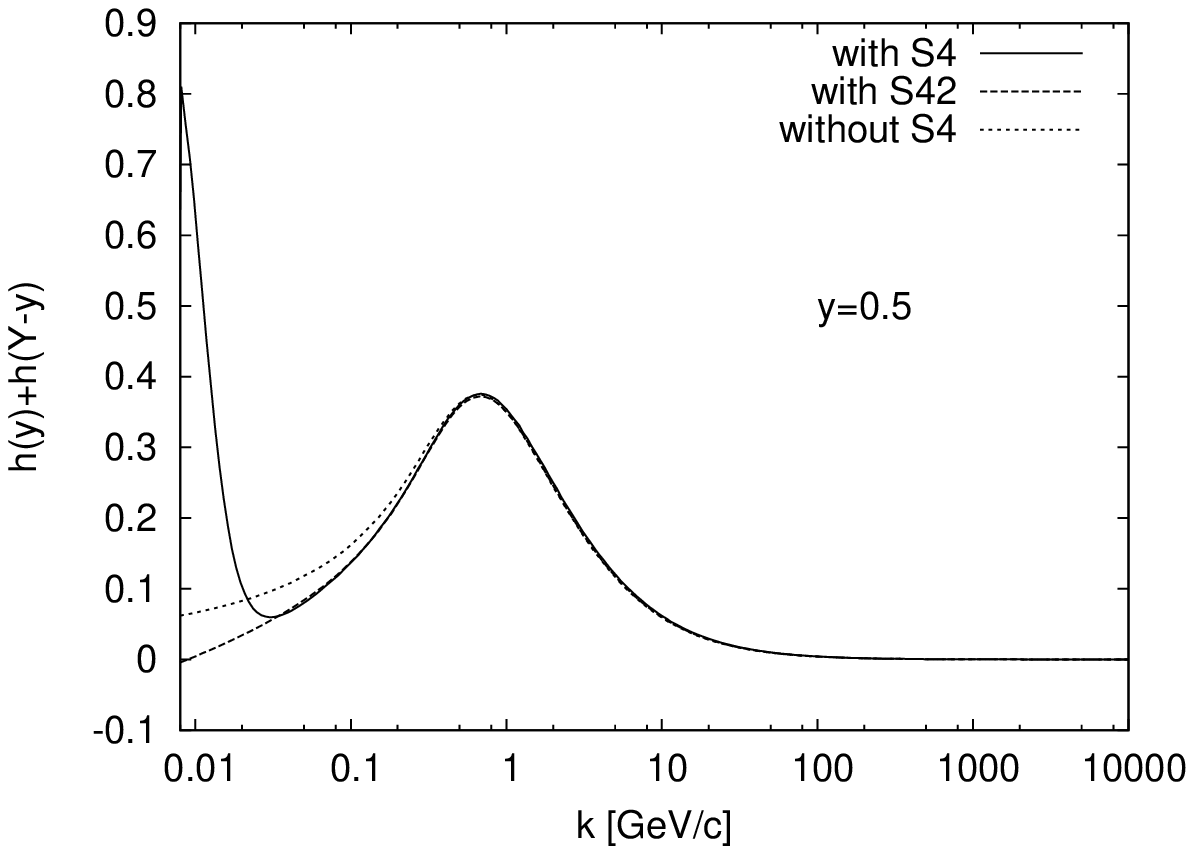}
\includegraphics[width=8 cm]{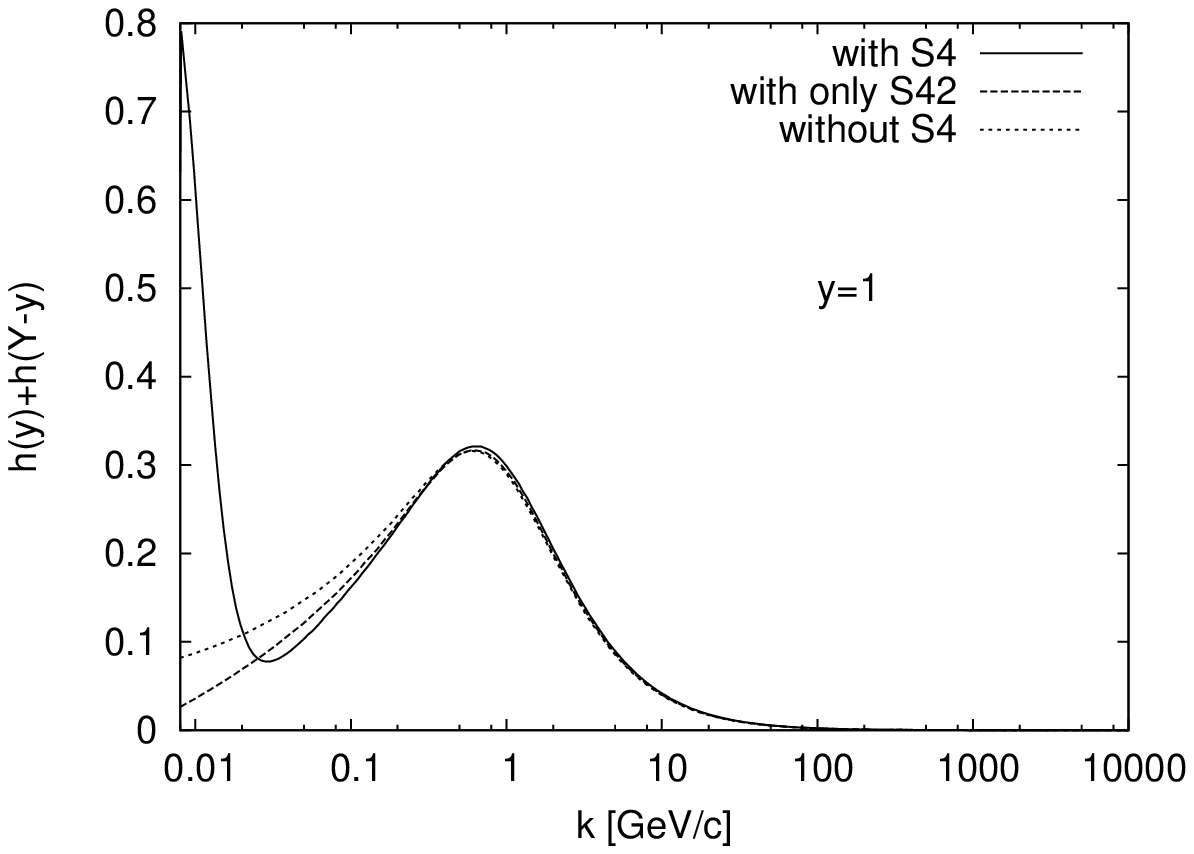}
\caption{Gluon densities for O-O collisions at $Y=2$ and $\bal=1$ with
and without  the quadruple pomeron interaction $S_4=S_4^{(1)}+S_4^{(2)}$ at intermediate rapidities $y=0,0.5$ and 1. }
\label{fig6}
\end{center}
\end{figure}

A more illustrative  comparison is done in Fig. \ref{fig6}.
One observes first of all that at small momenta the contribution from the derivative coupling $S_4^{(1)}$ blows up, which only exhibits the region of bad convergence.
One cannot trust  the obtained values of the gluon density in this boundary region. One rather expects the true density to behave roughly as with only $S_4^{(2)}$,
the contribution from $S_4^{(1)}$ only slightly lowering the density, as one can conclude from the figures at momenta immediately before the peak. At the physically important
momenta  in the region of the peak itself and
 higher the contribution from the quadruple interaction turned out to be  very small due to the assumed high value of $\bal$.

\begin{figure}
\begin{center}
\includegraphics[width=8 cm]{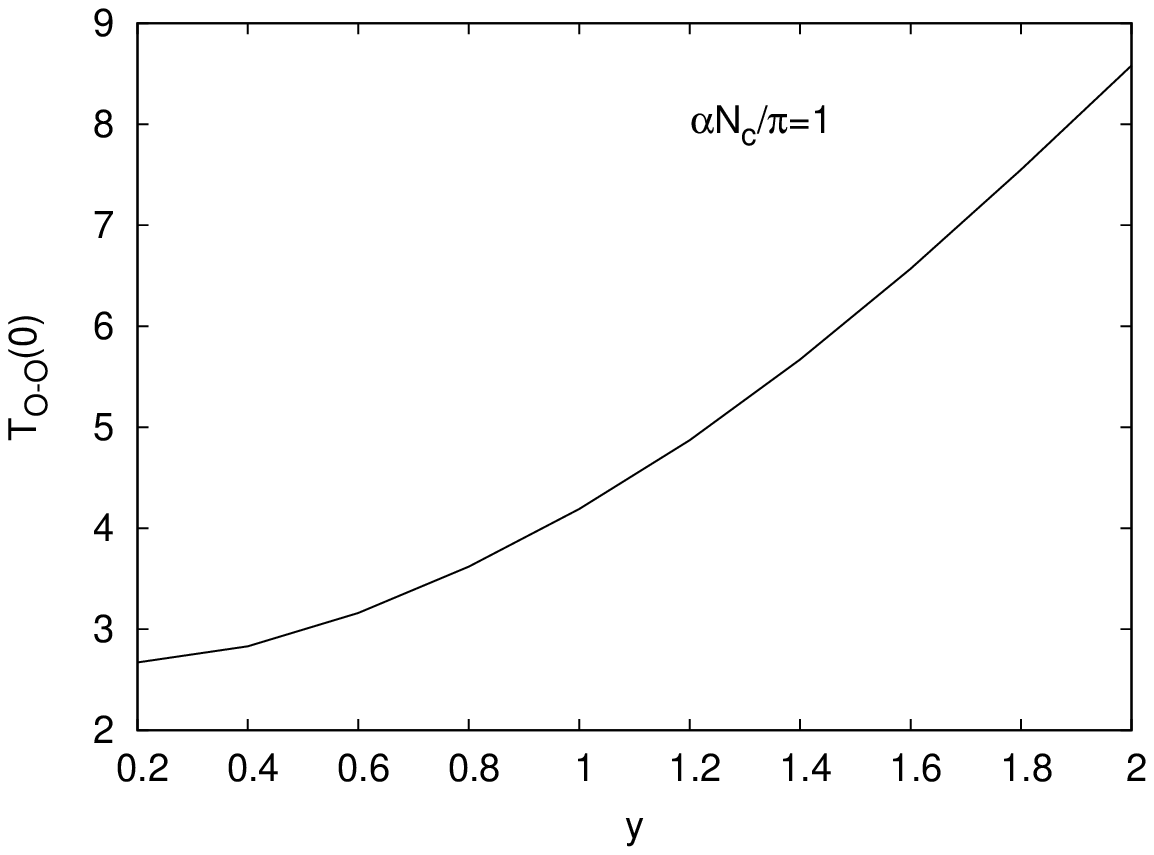}
\caption{Eikonal $T(0)$ for central O-O collisions at $Y=2$ with
 the quadruple pomeron interaction $S_4$ }
\label{fig7}
\end{center}
\end{figure}

The last conclusion is confirmed by the calculation of the eikonal function.
In Fig \ref{fig7} we present the eikonal  at ${\bf b}=0$ for $Y=2$,
which is obtained from the full action (\ref{clt1}) with the solution of  equations (\ref{cleq1n}) and (\ref{cleq2n})after integration over all $b_A=b_B$.
The difference between the present and absent
$S_4$ is smaller than 1\% (and cannot be shown in the plot).

To conclude with the iterative solution of the quasi-classical equations for the fields, we kave found that  it  turns out to be
dangerous in the infrared region. Probably the equations, being highly non-linear  possess solutions of a different type,
which avoid these difficulties but cannot be found by iterations.

\begin{figure}
\begin{center}
\includegraphics[width=8 cm]{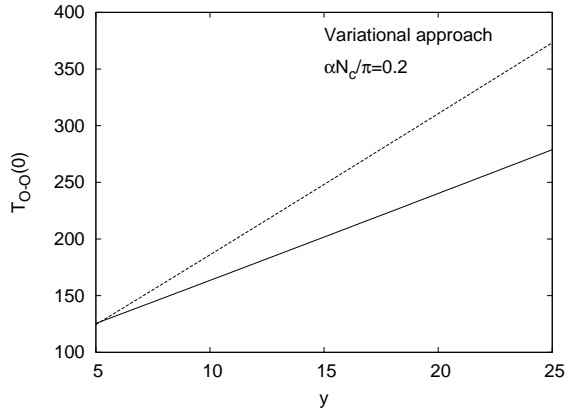}
\caption{Eikonal $T(0)$ for central O-O collisions obtained by the variational approach,
with the quadruple pomeron interaction $S_4$ (lower curve) and without it (upper curve)}
\label{fig8}
\end{center}
\end{figure}

Passing to direct variational search for the minimum of the action, we recall that without the
quadruple interaction one always found the minimum using the trying function (\ref{tryf})
with a variable $\Delta$.  Now we repeat this exercise with the quadruple interaction included.
We drop the infrared cut and take a more reasonable value of $\bal=0.2$, which of course substantially enhances the quadruple interaction.
We find that inclusion of $S_4$ allows to find the minimum of the action up to rescaled rapidities 5 that is
up to natural rapidities 25 but not to higher rapidities. So $S_4$ leads to certain restriction on the rapidity range,
which however is physically not so bad. To simplify calculations we choose to calculate the action
in the approximation of a constant $T_A(b_A)$ for each nucleus and afterwards obtained the eikonal function T(b)
integrating the action in the overlap area
of the collision depending on the impact parameter $b$. The obtained eikonal for central O-O collisions $T(0)$ is
illustrated in Fig \ref{fig8} (lower curve). To compare we also present $T(0)$ for the case without $S_4$ (upper curve).
 One observes that inclusion of the quadruple interaction substantially reduces the central eikonal. Still the eikonal remains
 very large numerically. As a result the total O-O cross-section is practically the same with or without $S_4$
 and with our parametrization equals 151 fm$^2$.

\section{Conclusion}
We have derived the quadruple pomeron interaction in the lowest order, which is essential in the interaction of nuclei between themselves.
 It consists of two terms, one with the derivative in rapidity and the other with the BFKL
interaction between the pair of pomerons. Both are infrared safe. Their  relative order to the triple pomeron interaction is $1/(\alpha_sN^2)$.
So the quadruple interaction is much smaller for fixed $\alpha_s$
and large number of colors. However in the opposite case with fixed $N_c$ and small coupling constant it may become comparable or even greater
than the currently used triple interaction. Illustrative and very approximate calculations of the O-O cross-sections  with the quadruple interaction together with the triple one
and quite large $\bar{\alpha}=1$ showed instability of the iterative procedure in the infrared region but a workable variational approach.
We find a noticeable damping of the gluon density at small momenta, although at momenta greater than 1 GeV/c the change is quite small.
In the variational approach one finds quite a substantial fall of the eikonal function at all impact parameters due to the quadruple interaction.
More precise calculations of the AA collisions are needed to make a final conclusion of the importance of the quadruple interaction for the actual processes.

\section{Appendix. Some details of the calculation}

We  use the logarithmic variable $t$ putting
$k=\exp (t)$
where $k$ is measured in units provided by the initial condition
$ \phi(y,k)_{y=0}=\phi_0(k)$.
Integration over $k$ transforms as
\[\int\frac{d^2k}{(2\pi)^2}=\frac{1}{2\pi}\int_{t_{min}}^{t_{max}}k^2dt.\]
The integral was done by the Simpson numerical integration in $n=400$ points.
Without infrared cut satisfactory results were achieved for
$t_{min}=-20$, $t_{max}=+20$. The infrared cut needed for iterations with $S_4^{(1)}$ was taken $t_{min}=-5$.

\subsection{Equation}
The equation to solve can be rewritten in the form
\beq\frac{\pd\phi}{\pd y}=+T_0+T_3^{(1)}+T_3^{2)}++T_4^{(1)}+T_{4}^{(2)}+T_E,
\label{cleq1a}
\eeq
where the common arguments are $(y,k)$ and terms $T_0$, $T_3^{(1,2)}$, $T_4^{(1,2)}$ and $T_4$
come from the parts $S_0$, $S_3$, $S_4$ and $S_E$ of the action.
Suppressing the common argument $y$ where it is possible, we have
\beq
T_0(k)=-\int \frac{d^2k_1}{4\pi^2}H(k,k_1|k,k_1)\phi(k_1).\eeq
After angular integration and passing to the integration variable $t_1$
\beq
T_0(k)=\bal\int dt_1 \Big(B(k,k_1)\phi(k_1)-A(k,k_1)\phi(k)\Big),
\label{t0}
\eeq
where
\beq
B(k,k_1=2\frac{k_1^2}{|k^2-k_1^2|},\ \ A(k,k_1)=2k^2
\Big(\frac{1}{|k^2-k_1^2|}-\frac{1}{\sqrt{k^2+4k_1^2}}\Big).
\label{kernab}
\eeq

From $S_3$ we have two terms
\beq T_3^{(1)}(k)=-\frac{2\alpha_s^2N_c}{\pi}\phi^2(k),\eeq
\beq T_3^{(2)}(k)=\frac{4\alpha_s^2N_c}{\pi}K^{-1}\phid(k)K\phi(k).\eeq

Operator $K^{-1}$ is nonlocal with the kernel
\cite{braun1,braun}
\beq
K^{-1}(k,k_1)=\frac{\pi}{2}\frac{1}{k_>^2}\Big(\ln\frac{k_>}{k_,}+1\Big)
\label{km1}
\eeq
where $k_>(k_<)$ is the greater(smaller) of momenta $k,k_1$.

So after angular integration we get
\[
T_3^{(2)}(k)=-\frac{2\alpha_s^2N_c}{\pi}
\Big\{\int_{-\infty}{t} dt_1 e^{-2z}\phid_2(k_1)
\Big((z+1)\phi_2(k_1)-2\phi_1(k_1)\Big)\]\beq+\int_{t}^{+\infty}dt_1\phid_2(k_1)\Big((1-z)\phi_2(k_1)
+2(2z-1)\phi_1(k_1)-4z\phi(k_1)\Big)\Big\},
\label{t32}
\eeq
where $z=t-t_1$, $\phi_1=(\pd/\pd t)\phi$, $\phi_2=(\pd/\pd t)^2\phi$.

The two quadruple interaction terms are
\beq
T_4^{(1,2)}(k)=-\frac{2}{N_c^2}\int\frac{d^2k_1}{(2\pi)^2}K^{-1}(k,k_1)\chi^{(1,2)}(k_1),
\label{t41}
\eeq
where
\beq
\chi^{(1)}(k)=\phid(k)\phi(k)\pd_y\phi(k),
\label{chi1}
\eeq
and
\beq \chi^{(2)}(k)=\phi(k)\int \frac{d^2k_1}{4\pi^2}\phid(k_1)H(k,k_1|k_1,k)\phi(k_1)
\label{chi2}
\eeq
Doing the angular integration and passing to integration variable $t_1$ we obtain, similarly to
(\ref{t0}):
\beq
\chi^{(2)}(k)=-\bal\phi(k)
\int dt_1 \Big(B_1(k,k_1)\phi(k_1)\phid(k_1)-A(k,k_1)\phi(k)\phid(k)\Big),
\label{chi}
\eeq
where $A$ is the same as in (\ref{kernab}) and
\beq
B_1(k,k_1)=\frac{k^4+k_1^4}{k^2}.
\label{b1}
\eeq

Finally the external term is
\beq
T_E(y,k)=\frac{1}{2}K^{-1}w_A(k)\delta(y).
\eeq

To obtain our final expressions for calculation we first rescale $\phi$, $\phid$ and $y$ as indicated in (\ref{rescale})
and afterwards divide by $2\alpha_s\bal$. Then the coefficients in $T_0$ and $T_3^{(1)}$ turn to unity
and in $T_3^{(2)}$ to 1/2.
The coefficient in $T_4$ is divided by $4\alpha_s^2 \bal$. so we get for the total coefficient
$1/(2\pi^2\bal^2)$.
Finally the rescaling of $y$ produces factor $\bal$ in $T_E$  so that the
 coefficient becomes  equal to $\alpha_s$.

As a result after rescaling equation (\ref{cleq1}) takes the form
(\ref{cleq1a}) with rescaled $y$ and $T_0,...T_E$ which are explicitly
\beq
T_0(k)=\int dt_1 \Big(B(k,k_1)\phi(k_1)+A(k,k_1)\phi(k)\Big),
\label{t0r}
\eeq
\beq T_3^{(1)}(k)=-\phi^2(k),
\label{t31r}
\eeq
\[
T_3^{(2)}(k)=-
\frac{1}{2}\Big\{\int_{-\infty}^{t} dt_1 e^{-2z}\phid_2(k_1)
\Big((z+1)\phi_2(k_1)-2\phi_1(k_1)\Big)\]\beq+\int_{t}^{+\infty}dt_1\phid_2{k_1)\Big((1-z)\phi_2(k_1)
+2(2z-1)\phi_1(k_1}-4z\phi(k_1)\Big)\Big\},
\label{t32r}
\eeq
where $z=t-t_1$,
\beq
T_4^{(1,2)}(k)=\frac{1}{2\pi^2\bal^2}\,\frac{1}{4}
\int dt_1k_1^2\frac{1}{k_>^2}(t_>-t_<+1)\chi^{(1,2)}(k_1),
\label{t4r}
\eeq
where $k>,t>$ and $k<,t<$ are the larger and small of $k,t$ and $k_1,t_1$ and the rescaled $\chi$ are
given by (\ref{chi1}) and (\ref{chi2}) without coefficient $\bal$. Finally
\beq
T_E(y,k)=\frac{\pi}{N_c}K^{-1}w_A(k)\delta(\bar{y}).
\label{ter}
\eeq

From the equation  at $y=0$ it follows that $\phi(y=0,k)=\alpha_sK^{-1}w_A(k)$, so that
\beq
\frac{1}{2}K^{-1}w_A(k)=\frac{1}{2\alpha_s}\phi(y=0,k),\ \ {\rm or}\ \
w_A(k)=\frac{1}{\alpha_s}K\phi(y=0,k).
\label{phi0}
\eeq

\subsection{Action}
The action is given by Eqs. (\ref{s0a}), (\ref{s3a}), (\ref{s41}), (\ref{s42a}) and (\ref{sexta}) and is a sum
\beq
S=S_0+S_3+S_4+S_E
\label{sa}
\eeq
Here we present it in the explicit form.

We begin with $S_0$ which is given by (\ref{s0a}). It contains two terms: one with the derivative $S_0^{(1)}$ and the second $S_0^{(2)}$ with $H$.
We split  operator $K$ into the product $K=L^\dagger L$.
In the logarithmic variable $t$ one finds $L=\pd^2/\pd t^2$. So after the angular integration  we get the part $S_0^{(1)}$
\beq
S_0^{(1)}=\frac{1}{2\pi}\int_0^Y dy \int dt k^2\Big(\phid_2(y,k)\frac{\pd\phi_2(y,k)}{\pd y}-\phi_2(y,k)\frac{\pd \phid(y,k)}{\pd y}\Big).
\label{s01}
\eeq
where we again denoted $\phi_2=\pd^2\phi/\pd t^2$. The integration over $y$ is originally performed
in the region $-\infty<y<+\infty$ and so takes into account the jumps at $y-0$ and $y=Y$. Contribution from these jumps gives the boundary term
\beq
S_0^{(b)}=\frac{1}{\pi} \int dt k^2\phid_2(0,k)\phi_2(0,k).
\label{s0b}
\eeq
where we have taken into account that $\phid(0)=\phi(Y)$,
The second term is
\beq
S_0^{(2)}=\frac{\bal}{\pi}\int dt dt_1 k^2\phid_2(y,k)
\Big(B(k,k_1)\phi_2(y,k_1)+A(k,k_1)\phi_2(y,k)\Big).
\label{s02}
\eeq
Here and in the following $k_1=\exp t_1$.

In $S_3$ both terms give the same contributions due to the assumed relation $\phid(y)=\phi(Y-y)$
So
\[
S_3=
\frac{8\alpha_s^2N_c}{\pi}\int dy\frac{d^2k}{4\pi^2}L\phid(y,k)L\phi^2(y,k)\]\beq=
\frac{8\alpha_s^2N_c}{\pi^2}\int dy dt k^2\phid_2(y,k)\Big(\phi_2(y,k)\phi(y,k)+\phi_1^2(y,k)\Big).
\label{s3b}
\eeq

Next
\beq
S_4^{(1)}=\frac{1}{\pi N_c^2}\int dy\int  dt k^2{\phid}^2\phi\pd_y\phi.
\label{s41ap}
\eeq
Presence of the derivative $\pd_y$ leads to the contribution from the jumps at $y=0$ and $y=Y$ and generates
 a new boundary term
\beq
S_4^{(b)}=\frac{1}{2\pi N_c^2}\int dt k^2{\phid}^2(0,k)\phi^2(0,k).
\label{s4b}
\eeq
The second part is
\beq
S_4^{(2)}=\frac{\bal}{\pi N_C^2}\int dy dt dt_1k^2\phi(k)\phid(k)\Big(B_1(k,k_1)\phi(k_1\phid(k_1)-A(k,k_1)\phi(y,k)\phid(y,k)\Big).
\label{s42}
\eeq
Finally
\beq
S_E=\frac{1}{\pi}\int dt k^2 w_A(k)\phid(0,k),
\label{seb}
\eeq
where $w_A$ is given by (\ref{phi0}).


\begin{thebibliography}{100}
\bibitem{jimwlk}J.~Jalilian-Marian, A.~Kovner, A.~Leonidov and  H.~Weigert
Nucl.\ Phys.\ B{\bf 504} (1997) 415; Phys.\ Rev.\ D {\bf 59} (1999) 014014;
E.~Iancu, A.~Leonidov, L.~McLerran, Nucl.\ Phys.\ A {\bf 692} (2001) 583;
E. Iancu, A. Leonidov, L. McLerran, Phys.\ Lett.\ B {\bf 510} (2001) 133;
E.~Ferreiro, E.~Iancu, A.~Leonidov and L.~McLerran, Nucl.\ Phys.\ A {\bf 703}
(2002) 489;  H.~Weigert, Nucl.\ Phys.\ A {\bf 703} (2002) 823.
\bibitem{lipatov} L.N.Lipatov, Nucl. Phys. {\bf B 452} (1995) 369; Phys. Rep., {\bf 286} (1997) 131
\bibitem{bal}I.~Balitsky,  Nucl.\ Phys.\ B {\bf 463} (1996) 99.
\bibitem{kov} Y.~V.~Kovchegov,Phys.\ Rev.\ D {\bf 60} (1999) 034008;{\it ibid}{\bf 61} (2000) 074018.
\bibitem{braun1} M.A,Braun, Phys. lett. {\bf B 483} (2000) 115
\bibitem{braun2} M.A.Braun, Phys. Lett. {\bf B 632} (2006) 297
\bibitem{braun3} M.A.Braun. Eur. Phys. J. {\bf C 48} (2006) 511
\bibitem{braun} M.A.Braun, Eur. Phys. J. {\bf C 33} (2004) 113
\bibitem{bond1} S. Bondarenko, L.Motyka,Phys. Rev, {\bf D 75} (2007) 114015
\bibitem{bond2} S. Bondarenko, M.A.Braun, Nucl. Phys. {\bf A 799} (2008) 151
\bibitem{dusling} K.Dusling, F. Gelis, T. Lappi, R. Venugopalan, Nucl Phys. {\bf A 836} (2010) 159.
\bibitem{mueller} S. Bondarenko, L. Motyka, A.H.Mueller, A.I.Shoshi, B.W. Xiao, Eur. Phys. J. {\bf C 50} (2007) 593.
\bibitem{braun4} M.A. Braun, G.P.Vacca,  Eur. Phys. J. {\bf C 50} (2007) 857.
\bibitem{braun5} M.A.Braun, Eur. Phys. J. {\bf C 73} (2013): 2418
\bibitem{BLV} J.Bartels, L.N.Lipatov, G.P.Vacca, Nucl. phys, {\bf 706} (2005) 391
\bibitem {braun6} M.A.Braun, Eur. Phys. J. {\bf C 70} (2010) 73
\end{thebibliography}
\end{document}